\documentclass[%
  reprint,
  color,
 amsmath,amssymb,
 aps,
 prd,
 twocolumn,
 float,
 longbibliography,
 nofootinbib
]{revtex4-2}

\usepackage{rotating}
\usepackage{graphicx}
\usepackage{dcolumn}
\usepackage{bm}
\usepackage{empheq}

\newcommand{\const}{{\rm const.}}

\newcommand{\sign}{\,{\rm Sign}}

\newcommand{\wde}{w_{\rm \Delta}}
\newcommand{\rmin}{r_{\rm min}}
\newcommand{\sep}{{\cal S}_{\Delta}}

\newcommand{\rplus}{r_+}
\newcommand{\rmoins}{r_-}

\newcommand{\mnras}{Mon. Not. of the Royal Astr. Society}
\newcommand{\aap}{Astron. \& Astrophys.}
\newcommand{\apss}{Astrophys. \& Space Science}

\newcommand{\apjs}{Astrophys. J. Suppl. Series}

\newcommand{\sovast}{Sov. Astron.}
\newcommand{\actaa}{Acta Astronomica}
\newcommand{\aj}{Astronomical J.}

\begin{document}

\title{Hill's level surfaces in the circular restricted three-body problem solved}

\author{Jean-Marc Hur\'e}

\affiliation{Universit\'e de Bordeaux, CNRS, LAB, UMR 5804, F-33600 Pessac, France}

\date{\today}

\begin{abstract}
We report the closed-form expression for Hill's surfaces in the circular restricted three-body problem. The solution $\phi(r,\theta)$, derived in the primary-centric spherical coordinate system, is deduced from a cubic equation delivering at most two roots on each side of a separatrix. The famous patterns (tadpole, horseshoe and peanut shapes, Roche lobes and Hill's quasispheres) are exactly produced.
\end{abstract}


\maketitle

\begin{figure}[h]
       \centering
       \includegraphics[width=8.cm,trim={-0.5cm 0cm 0.cm -0.5cm},clip]{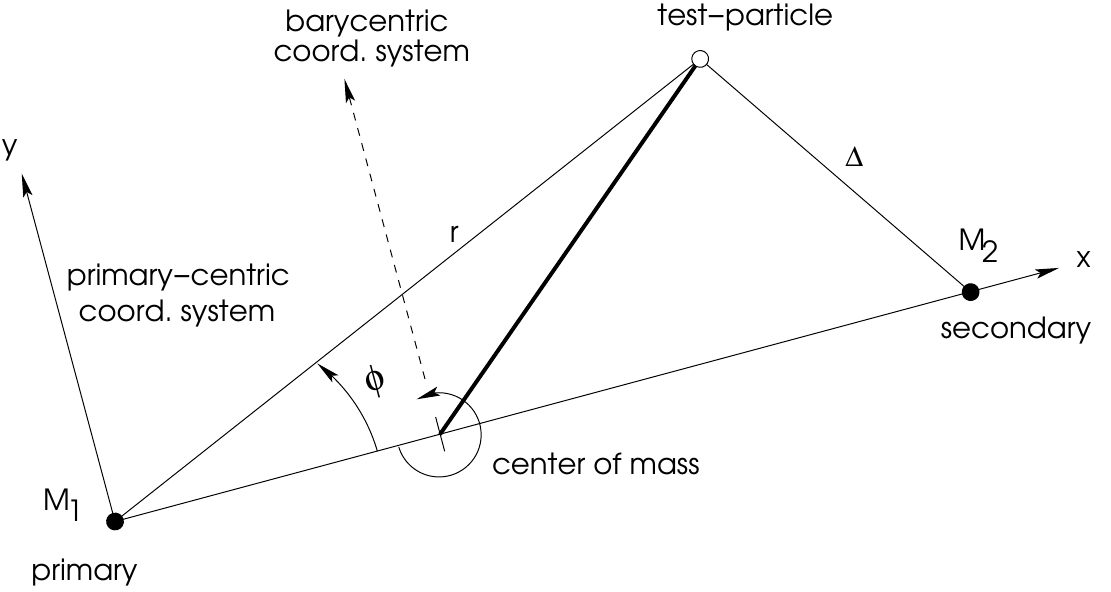}
       \caption{The primary-centric coordinate system attached to the synodic reference frame.}
       \label{fig:refs.pdf}
\end{figure}

\section{Context and challenges}

In the circular restricted three-body problem (hereafter CRTBP), a particle  moves in the gravity field of two pointlike massive stars orbiting each other at constant rate and separation. This problem has occupied a central place in celestial mechanics and dynamical astronomy for more than two centuries. It is fundamental for the physics of binary stars (e.g., Roche lobe overflow, light curves), for the dynamics of particles in star-planet systems and protoplanetary discs, and for space guidance and Solar System exploration. It is widely reviewed in textbooks (e.g.,\citep{s82,Bruno94,Murray_Dermott_2000,sf2aroche07,ak19}), and has several interesting varieties (e.g.,\citep{bc77,m78,s80,p95,c98,k09,djsf09,s11,agst12}). A remarkable property of the CRTBP is that the test particle is allowed to visit not all but certain regions of space, depending on the Jacobi constant $C = -2 W({\mathbf r}) - {\mathbf v}^2$, which is set by initial conditions in velocity ${\mathbf v}$ and position ${\mathbf r}$. The knowledge of equipotential surfaces  $W({\mathbf r})= \const$ (also known as Hill's surfaces or even ``zero-velocity surfaces'') that precisely separate permitted and forbidden regions is, therefore, of great interest. In the plane of the binary, these frontiers draw famous patterns: tadpole-shaped curves around equilateral Lagrange points, degenerating in horseshoe shapes as the potential decreases, then in a central peanut figure and an outer oval. Then comes the Roche lobes and Hill's quasispheres \citep{e83,ls07}. The literature apparently does not evoke the eventuality that $W({\mathbf r})$ is reversible, except some approximations \citep{s63,pm12}. Usually, an equipotential $W_0$ is found from three different numerical techniques, which are easily accessible with modern programming language and software: i) by successive approximations based on the fixed-point interaction, for instance $z=f(z;x,y,W_0)$, ii) by integration of a total derivative formed by differentiating $W$, e.g. $dx/dy|_{w_0}= -\partial_y W/\partial_x W$ in a plane $z = \const$, or iii) by detecting the zeros of the function $W(x,y,z)-W_0$ inside a precomputed grid from a root-finding method. In this article, we derive the solution ${\mathbf r}(W_0)$ by analytical means, therefore avoiding any systematic and complicated numerical procedure, and the generation of tables (e.g., \citep{pk64,m84}). The solution is exact, generally degenerate (roots of a cubic equation), and involves no series but standard functions. 

\section{Level surfaces from a cubic}
\label{sec:cubic}


In the CRTBP, the coordinate system is usually rotating with the two bodies. We follow this convention: the $x$-axis is chosen as the binary axis, and $z$ is perpendicular to the binary's plane of motion, as depicted in Fig. \ref{fig:refs.pdf}. With the standard barycentric frame, the distance between this particle and each star depends explicitly on the mass ratio, which is not optimal here. Instead, we use the primary-centric coordinate system, where one of the stars (the primary, here) stands at the origin (e.g.,\cite{s23}). If all lengths are normalized to the binary separation $a$, then the potential $w$ that dictates the displacement of the particle at $\bm{r}(x,y,z)$, in units of $GM_1/a$, is
\begin{equation}
w(x,y,z) = -\frac{1}{r}-\frac{q}{\Delta}-\frac{1+q}{2}\left[\left(x-\frac{q}{1+q}\right)^2+y^2\right],
\label{eq:w}
\end{equation}
where $q=M_2/M_1 \in \, ]0,1]$ is the mass ratio, and $\Delta$ is the distance from the secondary (with mass $M_2$), with
\begin{equation}
  \Delta^2 = (x-1)^2+y^2+z^2.
      \label{eq:bigdelta}
\end{equation}

\begin{figure*}
       \centering
       \includegraphics[width=18.1cm,trim={0.9cm 12cm 0.cm 0.5cm},clip]{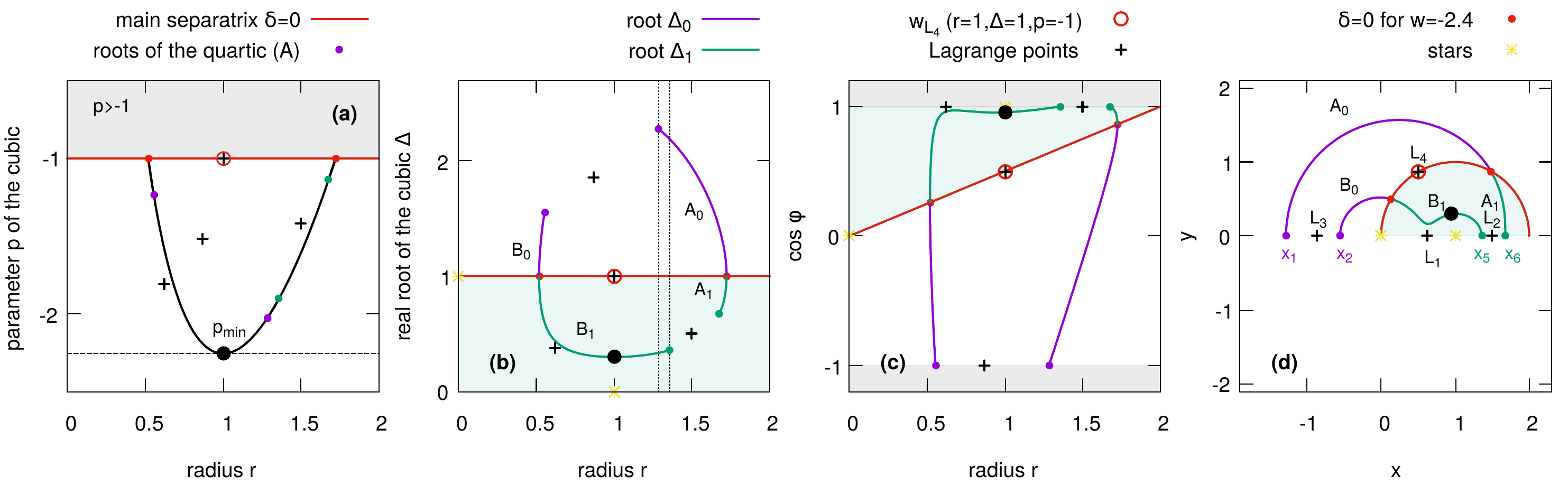}
       \caption{Values of the parameter $p < -1$ of the depressed cubic  ({\it left}), computed from \eqref{eq:defp} by scanning the range $r \in [0,2]$, for $q=0.3$, $\sin \theta=1$ and $w_0=-2.4$ (case III, see Sec. \ref{sec:computing}). Positive roots of the cubic versus the radius ({\it middle left}). 
         For $1.28 \lesssim r \lesssim 1.36$ ({\it vertical lines}), branches A$_0$ and B$_1$ are fed simultaneously. Cosine of the azimuth $\phi$ deduced from \eqref{eq:phidelta} ({\it middle right}). Equipotential (oval and interior peanut shapes) in the  plane of the binary ({\it right}), limited to the half-plane $y\ge 0$.}
       \label{fig:w24q03test.pdf}
\end{figure*}

By converting \eqref{eq:w} and \eqref{eq:bigdelta} in spherical coordinates $(r,\theta,\phi)$ still attached to the primary-centric frame (see Fig. \ref{fig:refs.pdf}), we notice that the azimuth $\phi$ can be eliminated in the first equation, leading to a {\it depressed cubic equation} in the variable $\Delta$, namely
    \begin{equation}
      \Delta^3 + 3 p \Delta + 2s = 0,
      \label{eq:cubic}
    \end{equation}
    
    where $s=1$, and
    \begin{flalign}
     &3 p = \frac{2}{q}w+ \frac{2}{qr}-\frac{1}{1+q} + r^2\left(\frac{1+q}{q} \sin^2 \theta -1 \right).
     \label{eq:defp}
    \end{flalign}
    This simple manipulation offers a direct way to construct a given equipotential $w(x,y,z)=w_0$ because the cubic equations are solvable by analytical means. There are four steps : i) selection of a pair of coordinates $(r,\theta)$, ii) determination of the parameter $p \equiv p(q,w_0;r,\theta)$ from \eqref{eq:defp}, iii) calculation of the roots of the depressed cubic (if any; see below), and iv) calculation of the coordinate $\phi$ through its cosine from \eqref{eq:bigdelta}, that is
    \begin{equation}
      \cos \phi = \frac{1+r^2-\Delta^2}{2 r \sin \theta} \in [-1,1].
      \label{eq:phidelta}
    \end{equation}

A level curve is, therefore, found by varying either the radius $r \in ]0,\infty[$ or the colatitude angle $\theta \in ]0,\pi[$, or both, then leading to surface levels\footnote{Note that with the function $w$ being symmetric with respect to the $xz$- and $xy$-planes, the analysis can be limited to $y \ge 0$ and $z \ge 0$.}. 

      \subsection{Real, positive roots}

      The nature of the three roots of \eqref{eq:cubic} depends on the sign of the discriminant $\delta=-108(1+p^3)$. If $p>-1$ there are two complex roots and a real root, but it can be shown that $\Delta < 0$ in this case, which is not acceptable. The three real roots obtained for $p \le -1$ are
        \begin{equation}
\Delta_k= 2 \sqrt{-p} \cos \left[\frac{1}{3} \arccos \left(\frac{s}{p\sqrt{-p}}\right) - \frac{2 k \pi}{3} \right],\label{eq:xk012}
        \end{equation}
        with $k \in \{0,1,2\}$, but only $\Delta_0$ and $\Delta_1$ are positive and relevant. Then, two cosine $\cos \phi_k$ follow from \eqref{eq:phidelta}, which is enough to get the points on the equipotential.

\subsection{The main separatrix and the second cubic}
\label{sec:circularseparatrix}

The case $p=-1$ corresponds to a vanishing discriminant $\delta$ for the cubic, and we have $\Delta_0=\Delta_1=1$. We see from \eqref{eq:defp} and \eqref{eq:phidelta} that it corresponds to a certain surface, a {\it separatrix}, denoted $\sep$, which separates the space into two regions \citep{matas78}. From \eqref{eq:phidelta}, we get  $x=r \sin \theta \cos \phi=r^2/2$ (then, $y$ and $z$ easily follow), meaning that this separatrix is the sphere with unit radius and the secondary as the center. It can be shown that the relevant root is $\Delta_0$ (and $\Delta_1$)  when ${\mathbf r}$ is outside (inside, respectively) $\sep$. Setting $p=-1$ in \eqref{eq:defp} leads to another depressed cubic
    \begin{equation}
      r^3 + 3 p' r + 2s' = 0
      \label{eq:bigr}
    \end{equation}
where
\begin{subequations}
    \begin{empheq}[left={\empheqlbrace}]{align}
      3p' & = s' \left[2w + q\left(\frac{3q+2}{q+1}\right)\right],\label{eq:defsprim}\\
s' = &\frac{1}{(1+q) \sin^2 \theta -q}.\label{eq:deftprim}
    \end{empheq}
\end{subequations}
As a consequence, the roots of this new cubic equation (if any) give the positions where $\Delta_0$ and $\Delta_1$ meet on $\sep$. Again, the real, positive roots of \eqref{eq:bigr}, denoted $r_k$, depend on the sign of the discriminant $-108({s'}^2+{p'}^3)$. As for the first cubic, and for the same reasons, only the roots obtained for $k \in \{0,1\}$ are relevant\footnote{Formally, the expression for the three roots $r_k$ is similar to \eqref{eq:xk012}.}. When ${p'}^3=-{s'}^2$, \eqref{eq:bigr} has a double root, i.e. $r_0=r_1=1$. This happens for a special value of $w_0$ that is deduced from \eqref{eq:defsprim} and \eqref{eq:deftprim} (it depends on $\theta$). As for the first cubic, there is a closed curve belonging to $\sep$, denoted ${\cal C}$ (another separatrix), where $r_0$ and $r_1$ merge. That curve is in fact the circle with radius $\sqrt{3}/2$ and center at Cartesian coordinates $(\frac{1}{2},0,0)$. It meets the plane of the binary at the triangular Lagrange point L$_4$ because $r=\Delta=1$, and we have
\begin{equation}
  w_{L_4}=-\frac{3q^2+5q+1}{2(1+q)}.
  \label{eq:wmax}
\end{equation}
We set $\rmoins \equiv r_1$ and $\rplus = r_0 \ge \rmoins$ from now on.

\begin{figure*}
       \centering
       \includegraphics[width=18.1cm,trim={0.9cm 12cm 0.cm 2.5cm},clip]{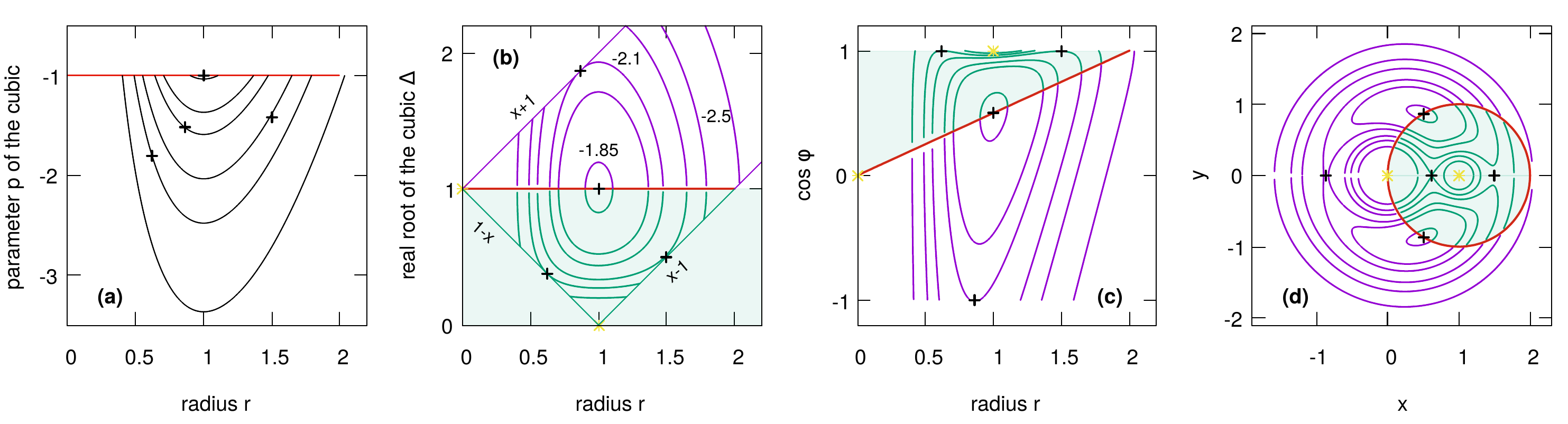}
       \caption{Same legend (simplified) as for Fig. \ref{fig:w24q03test.pdf} but for $w_0 \in \{1.85,-2\}$ (tadpole shape), $w_0 \in \{-2.1,-2.3\}$ (horseshoe shape), $w_0=-2.5$ (interior peanut shape), and $w_0=-2.9$ (Hill's quasispheres; three components).}
       \label{fig:fivewq03test.pdf}
\end{figure*}

\subsection{Equatorial plane of the binary}
      
      The orbital plane of the stars has always been of immense interest and the shape of equipotentials is relatively familiar (e.g., \citep{binneytremaine87}), see the Introduction. This is the reason why we propose to illustrate the method by setting $\theta=\frac{\pi}{2}$. An equipotential  at $z=0$ can be composed of at most three components depending on $w_0$ (by order of appearance as $w_0$ decreases):
      \begin{itemize}
      \item component A, for tadpole and horseshoe shapes, obtained for the highest permitted values of $w_0$,
      \item component B, for the interior peanut shape, for ``intermediate'' values of $w_0$ 
      \item component C, for the Roche lobes and Hill's quasispheres, drawn as  $w_0 \rightarrow -\infty$.
      \end{itemize}
      It follows from above that any component of the equipotential $w(x,y,0)=w_0$ that crosses the separatrix $\sep$ once is composed of two branches, because of the two roots for the cubic (this holds for $z>0$). For component A, one branch is computed with $\Delta_0$ (it is denoted A$_0$), and the other, denoted A$_1$, is obtained from $\Delta_1$, and so on for components B and C. Figure \ref{fig:w24q03test.pdf} shows a first example obtained with a mass ratio $q=0.3$ and for $w_0=-2.4$, by basically scanning the range $r \in [0,2]$ with regular spacing. In this interval, all values of $r$ do not lead to $p \le -1$ and subsequently to positive roots (any complex or negative roots are rejected). Panel (a) shows that $p \le -1$ only for $ 0.518 \lesssim r \lesssim 1.723$, (again, see Sec. \ref{sec:circularseparatrix}), and there is a minimum\footnote{This minimum $\rmin$ of $p$ is easily found by differentiation of \eqref{eq:defp}, which leads to $\rmin^3\left[(1+q) \sin^2 \theta -q\right]=1$.\label{fn1}} at $r=1$. The physical roots $\Delta_0$ and $\Delta_1$ of the cubic equation are displayed in panel (b), and $\cos \phi$ is plotted in panel (c). Panel (d) shows the components A and B of the equipotential in the $xy$-plane, deduced from the standard spherical to Cartesian conversion formula combined with \eqref{eq:phidelta}. Figure \ref{fig:fivewq03test.pdf} displays a set of six equipotentials computed in a similar fashion, again by a raw scan of the interval $r \in [0,2]$. We have selected  values of $w_0$ corresponding to the four classical patterns, from tadpole shape to Hill's quasispheres.
      
 \subsection{Remarks}

 This method yields exact solutions and works very well. Three slight issues must be quoted. First, the radial sampling plays a crucial role, especially when $-w$ becomes large. As expected, the equipotentials become more and more circular as the energy decreases (in the vicinity of each star, and far enough from the system), and $\cos \phi$ becomes less and less sensitive to $r$; see Fig. \ref{fig:w24q03test.pdf}c. This can be improved by a densification of points at the edges of branches (see below). A second difficulty is caused by a double degeneracy: a single value of the parameter$^{\ref{fn1}}$ $p \in [p(\rmin),-1]$ is obtained by two different values of $r$, and there are at most two roots at a given radius. It means that each branch is not revealed in a continuous fashion, but piecewise. This is visible in Fig. \ref{fig:w24q03test.pdf}(b): as $r$ decreases, the roots $\Delta_0$ and $\Delta_1$ feed both branches A$_0$ and A$_1$ of component A, then, there is a short range of radii for which both components A and B are constructed simultaneously (namely, $1.28 \lesssim r \lesssim 1.36$), then branches B$_0$ and B$_1$ are revealed in parallel. This can be circumvented by making an inventory of the intersections of an equipotential with the binary axis, as explained in Sec. \ref{sec:computing}. Third, unless a happy coincidence, there is no value of $r$ radius in the user-selected interval that exactly leads to $|\cos \phi| =1$ in \eqref{eq:phidelta}. It means that the intersections of the equipotential with the binary axis (if needed) are not expected to be output from the cubic. This point can be fixed (see Sec. \ref{sec:computing}).

\section{Computing branches individually. The equatorial plane as a fully solvable case}
\label{sec:computing}

\subsection{Preliminaries}
\label{sec:preliminaries}

We solve the cubic for a given mass ratio $q$ and level $w_0$ by decreasing or increasing the radius $r$ in a given interval, say $[r_a,r_b]$. As quoted already, the equipotential is revealed piece-by-piece and its components and branches simultaneously, because $r(p)$ is not a bijection and there are two roots for the cubic. If one wishes to isolate a branch in the form of a continuous curve, then we must determine in advance which intervals in radius are concerned. For this purpose, the following quantities are required:
\begin{itemize}
\item[i)] The position of the four Lagrange points ${\mathbf r}_{L_l}$, with $l \in [1,4]$, each one belonging to a specific equipotential, $w_{L_l}$, which is computed from \eqref{eq:w}.
\item[ii)] The intersections of the equipotential with the main separatrix $\sep$, i.e. the two real roots of the second cubic, $\rmoins$ and $\rplus \ge \rmoins$ of \eqref{eq:bigr}, see Sec. \ref{sec:circularseparatrix}. We find  $s'=1$ from \eqref{eq:deftprim}, and $w_0$ is given by \eqref{eq:wmax}.
\item[iii)] The intersections of the equipotential with the axis of the binary. We can establish that the equation that localizes these points is an ensemble of {\it three general quartic equations} (see Appendix \ref{app:intersection}). At most, there are six roots, denoted $x_k$ with $k \in \{1,6\}$, and sorted in increasing order $x_1 \le x_2 < x_3 \le x_4 < x_5 \le x_6$; see Fig. \ref{fig:w24q03test.pdf} where $x_1$, $x_2$, $x_5$ and $x_6$ are reported.
  
\item[iv)] The value of the equipotential, $\wde$, that crosses the binary axis at $r=2$ and $\Delta=1$ (and tangent to the main separatrix $\sep$). Actually, for $w \le \wde$, component A does not cross the separatrix anymore, and there is a single branch, namely, A$_1$. That value is directly deduced from \eqref{eq:defp} with $p=-1$,
\begin{equation}
 \wde = \frac{q^2+q-1}{2(1+q)} -2(1+q) \sin^2 \theta,
\end{equation}
which becomes $\wde=-\frac{3q^2+7q+5}{2(1+q)}$ for $\theta=\frac{\pi}{2}$.

\end{itemize}

\subsection{Procedure (decision tree)}

From these preliminary quantities, the procedure to construct a given component or branch is as follows (there is no solution for $w_0 > w_{L_4}$) :
\begin{itemize}
  
  \item Case I (tadpole shape): $w_{L_3} < w_0 \le w_{L_4}$. The equipotential has a single component A: branch A$_0$ (outside the main separatrix $\sep$) is obtained with $\Delta_0$ for $r \in [\rmoins,\rplus]$ (see Sec. \ref{sec:cubic}), while branch A$_1$ is obtained with $\Delta_1$ in the same interval in radius. Note that $\rmoins=\rplus=1$ at L$_4$.
  \item Case II (horseshoe shape): $w_{L_2} < w_0 \le w_{L_3}$. Again, there is single component. Branch A$_1$ is still revealed for $r \in [\rmoins,\rplus]$, but branch A$_0$ has two parts: the interior part is obtained for $r \in [-x_1,\rplus]$ and the exterior part is found for $r \in [-x_2,\rmoins]$. 

  \item Case III (interior peanut shape): $w_{L_1} < w_0 \le w_{L_2}$. The equipotential has two components, A and B. Component A has two branches, A$_0$ and A$_1$, computed for $r \in [-x_1,\rplus]$ and $r \in [\rplus,x_6]$, respectively. Component B also has two branches, B$_0$ and B$_1$, computed for $r \in [-x_2,\rmoins]$ and $r \in [\rmoins,x_5]$, respectively. Branch B$_0$ corresponds to the interior part of branch A$_0$ of case II.
    
  \item Case IV (Hill's quasispheres): $\wde < w_0 \le w_{L_1}$. The equipotential now has three components, A, B and C. Component A is determined as in case III. Component B surrounds the primary and is made of two branches, B$_0$ and B$_1$, computed for $r \in [-x_2,\rmoins]$ and $r \in [\rmoins,x_3]$, respectively. Component C has a single branch, C$_1$. It surrounds the secondary and is obtained for $r \in [x_4,x_5]$. Note that the Roche lobes appear for $w_0 = w_{L_1}$ when components B and C get in contact.
    
  \item Case V (Hill's quasi-spheres; continued): $w_0 \le \wde$. The three components are obtained as for case IV, except for component A, which is now fully outside the main separatrix. There is, therefore, a unique branch A$_0$, which is computed for $r \in [-x_1,x_6]$.
\end{itemize}

This five cases are easily implemented in any program. For each interval in radius here-above, the sampling in radius remains totally free (it can be uniform or not). We observe that the density of points obtained by a uniform spacing is generally low close to the edges of branches (i.e. vicinity of the main separatrix and vicinity of the binary axis). This can be corrected by an adaptive sampling, for instance from a scaled Chebyshev grid.

\begin{figure}
       \centering
       \includegraphics[width=7.5cm,trim={4.5cm 6cm 5cm 2cm},clip]{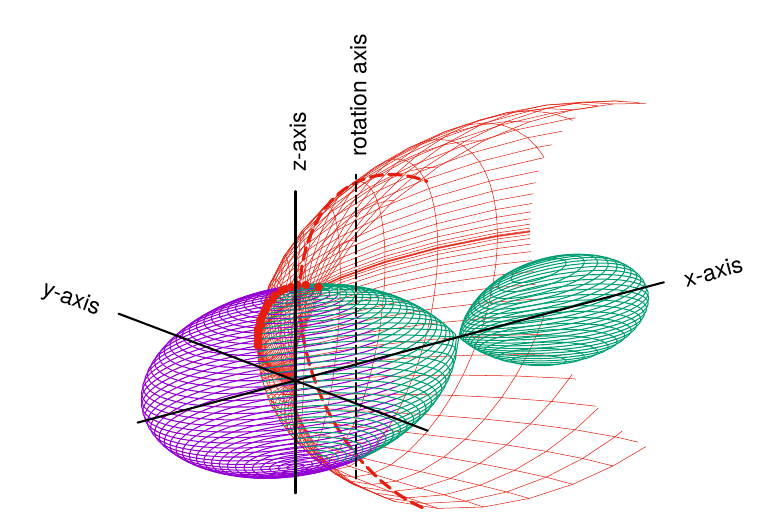}
       \caption{Components B and C of the equipotential $w_0=w_{L_1}$ in three dimensions (the Roche lobes) computed from the two roots $\Delta_0$ ({\it purple}) and $\Delta_1$ ({\it green}) of the cubic \eqref{eq:cubic} for a mass ratio $q=0.3$, the main separatrix $\sep$ ({\it plain red lines}) and the second separatrix ${\cal C}$ ({\it dashed red line}). The intersections at $\rmoins$ and $\rplus$, obtained from the second cubic \eqref{eq:bigr}, are also plotted ({\it red dots}).}
       \label{fig:rlobes03.pdf}
\end{figure}

\section{Conclusion}

We have shown that the equipotentials $W({\mathbf r})=\const$ (or, level surface of the Jacobi constant) in the CRTPB can be determined exactly by solving a cubic equation. 
This is possible by selecting one star (the primary) as the center of coordinates, leaving a single radical in the expression of the potential in the corotating frame, namely the separation from the secondary. Along the axis of the binary system, the equipotentials can be exactly located from quartics (this is also the case of the $z$-axis).
The function $W({\mathbf r})$ is, therefore, fully reversible at any point ${\mathbf r}$ of the equatorial plane of the binary (see Appendix \ref{sec:threedim} for the off-plane case). The formulae for the three components of ${\mathbf r}(W_0)$ are simple and easy to implement in codes. Numerical solvers (see the Introduction) can, therefore, be avoided to construct equipotential maps, which already represents a huge gain in terms of computing time. As usual in physics, having a formula is always a considerable advantage over numerical approaches, in particular when it is exact. This opens a few interesting perspectives, to begin with the dynamics of the test particle in the CRTBP itself. The expression  ${\mathbf r}(W_0)$ can be used for perturbative calculus, in the value of $W_0$, in the position ${\mathbf r}$ or in the mass ratio $q$. It is also interesting in the context of binary evolution, as long as the point-mass approximation for the two massive bodies is justified. The geometries of peanut shapes, Roche lobes and Hill's quasispheres rule mass transfers between the two components and subsequently impact the orbital elements \cite{p66,apa02,ak19,lfqb23,gh2024}. In particular, in semidetached and contact binaries, Roche lobe overflow is a complex problem which requires an accurate description of the equipotential in the vicinity of the inner Lagrange point L$_1$ where matter transits \citep{mpetal21,rsddvmjpr25}. The knowledge of a closed form for ${\mathbf r}(W_0)$ enables us to get the exact shape of the Roche lobes (see Fig. \ref{fig:rlobes03.pdf} as an example obtained with the formula with $q=0.3$) and around them, and this should help in improving the realism of models. Actually, the potential is usually Taylor expanded at point L$_1$ \citep{mpetal21,cp23}, which has limited validity, in contrast with the relationship ${\mathbf r}(W_0)$. The geometry of equipotentials is also a key ingredient for simulating light curves of close binaries \citep{a88,prsa16,cfp21}, from semidetached ot overcontact binaries. The results presented here are particularly well suited to such a task, as the local normal to any equipotential surface can be determined exactly by differential calculus from $\phi(r,\theta)$. This should considerably simplify the production of synthetic diagrams, that could serve as true references. 

\begin{acknowledgements}
I wish to thank H. Pelissard (LAB), and especially A. Pierens (LAB) and J.F. Bony (IMB) for valuable inputs on the paper before submission. The anonymous referee is acknowledged for suggestions to improve the paper.
\end{acknowledgements}

\bibliographystyle{apsrev4-1}
%

        \appendix


\section{Intersection with the axis of the binary}
\label{app:intersection}

Forcing $\cos \phi= \pm 1$ in \eqref{eq:phidelta} furnishes a relationship between $r$, $\theta$, $q$ and $w$, but we cannot use the depressed cubic to get $\Delta$. In order to describe the full $x$-axis, we set $r = \epsilon_1 x$ and $\Delta =  \epsilon_2(1- x)$, where $\epsilon_1 = \pm 1$ (sign $+$ is for $x>0$) and $\epsilon_2 = \pm 1$ (sign $+$ is for $x<1$). In these conditions, \eqref{eq:w} becomes
\begin{flalign}
  \nonumber
  &\left[w + \frac{q}{2(1+q)}\right] x(1-x) + \epsilon_1 (1-x) + q \epsilon_2 x  \\
  & \qquad+ \frac{1}{2}x^3(1-x) +\frac{1}{2}x(1-x)^3 =0.
\label{eq:wxaxis}
\end{flalign}
This is a general quartic in the variable $x$, which can be rewritten in the form $x^4+bx^3+cx^2+dx+e=0$, with
        \begin{subequations}
    \begin{empheq}[left={\empheqlbrace}]{align}
      &-b(1+q)=1+3q,\\
      &c(1+q)=2w- \frac{q}{1+q}+3q,\\
      &-d(1+q)=2w- \frac{q}{1+q}-2\epsilon_1+q(1+ 2\epsilon_2),\\
      &e(1+q)=-2\epsilon_1.
    \end{empheq}
        \end{subequations}
        There are in general four complex roots, but only real roots are physically acceptable here. By order of appearance, as $w < w_{L_4}$ decreases, we have the following:
\begin{itemize}
\item[$\bullet$] $x_1$ and $x_2>x_1$ are the two negative roots of the first quartic formed with $\epsilon_1=-\epsilon_2=-1$, located leftward to the primary on both sides of Lagrange point L$_3$. We have $x_1=x_2=-r_{L_3}$ at L$_3$, which marks the tadpole shape $\leftrightarrow$ horseshoe shape transition. 
\item[$\bullet$] $x_5$ and $x_6>x_5$ denote the two positive roots of the second quartic formed with $\epsilon_1=\epsilon_2=+1$ and located rightward to the secondary star, on both sides of Lagrange point L$_2$. We have $x_5=x_6=r_{L_2}$ at L$_2$, which corresponds to the horseshoe shape $\leftrightarrow$ peanut shape transition. 
\item[$\bullet$] $x_3$ and $x_4>x_3$ denote the two positive roots of the third quartic formed with $\epsilon_1=-\epsilon_2=+1$, and located on both sides of the inner Lagrange point L$_1$, in between the two stars. We have $x_3=x_4=r_{L_1}$ at L$_1$, which is the peanut shape $\leftrightarrow$ Hill's quasisphere transition. 
 \end{itemize}

\section{Off-plane solutions}
        \label{sec:threedim}
        
        By varying $\sin \theta$, one gets exact surface levels off the plane of the binary by the method explained in Sec. \ref{sec:cubic}, and without any modification. As an illustration, we show in Fig. \ref{fig:rlobes03.pdf} the two Roche lobes computed from the cubic equation for $q=0.3$ and $w_0=w_{L_1}$ and various colatitude angles $\theta$. The intersections of the equipotential with $\sep$, calculable from \eqref{eq:bigr}, the main separatrix $\sep$ and the curve ${\cal C}$ are also displayed. There is a slight shade between the equatorial case and the off-plane case $z>0$: the intersection of equipotentials with the $xz$-plane (if needed) no longer involves quartics, but higher-degree polynomials. As for the equatorial case, these intersections are not mandatory to make a map as the figure shows. However, it can be useful to get the off-plane analog of the $x_k$'s for some problems (see Sec. \ref{sec:preliminaries}). There are three cases:
        \begin{itemize}
         \item[a.] The intersection of an equipotential $w_0$ with the $z$-axis. It is found from \eqref{eq:w}, namely
  \begin{flalign}
          \nonumber
          &q^2-\left[w_0+\frac{q^2}{2(1+q)}+\frac{1}{z}\right]^2(1+z^2)=0,
  \end{flalign}
  which leads to another general quartic equation, in the variable $z\ge0$ (see Appendix \ref{app:intersection}). This case is, therefore, solvable by analytical means.
 \item[b.] The intersection of an equipotential with the main separatrix $\sep$. It is deduced from \eqref{eq:w} by solving
  \begin{flalign}
          \nonumber
          &w_0+\frac{1}{r}+q+\frac{1+q}{2}\left(\frac{r^2}{2}-\frac{q}{1+q}\right)^2=0,
  \end{flalign}
  which is a general quintic equation in the variable $r$. Note that, for $r=1$, the equipotential also belongs to ${\cal C}$ and this occurs for $w_0=-\frac{9+14q+9q^2}{8(1+q)}$. 
\item[c.] The intersection of an equipotential $w_0$ with the $xz$-plane (out of cases a and b). It involves an general octic equation, obtained by expanding
        \begin{flalign}
          \nonumber
          &\left[w_0 + \frac{1}{r}+\frac{1+q}{2}\left( r \sin \theta  \mp \frac{q}{1+q}\right)^2\right]^2\\
          & \qquad  \qquad \times \left(1+r^2 \mp 2 r \sin \theta \right)-q^2=0,
          \end{flalign}
        where $\pm = \cos \phi = \sign(x)$.
        \end{itemize}
        General quintics and higher-degree polynomial equations do not have roots involving radicals or standard functions \citep{King1996}. As a consequence, for cases b and c above, a numerical technique has to be employed to get these intersections (if needed), e.g. a root-finding method.\\
      
\end{document}